\documentclass[
    letterpaper, 
    10.5pt, 
    twoside, 
]{LTJournalArticle}

\newcommand{\keywords}[1]{%
  \vspace{0.5cm}
  \noindent\textbf{Keywords:} #1
  \vspace{0.5cm}
}

\usepackage[utf8]{inputenc} 
\usepackage{upgreek}
\usepackage{stfloats}
\usepackage{blindtext}
\usepackage{amsmath}
\usepackage{multirow}
\usepackage{physics}
\numberwithin{equation}{section}
\usepackage{times}
\usepackage{setspace}
\usepackage{import}
\usepackage{upgreek}
\usepackage{enumitem}
\usepackage{tabularx}
\usepackage{ragged2e}
\usepackage{lscape}
\usepackage{etoolbox}
\usepackage{lipsum}
\usepackage{url}
\usepackage{tablefootnote}
\usepackage{abstract}
\usepackage{multicol}
\usepackage{kotex}
\usepackage{epstopdf}
\usepackage[caption=false]{subfig}
\usepackage{soul}
\usepackage{authblk}

\title{Machine learning for characterizing uncertain elastic properties of fused filament fabricated materials for topology optimization applications}

\author[1]{Zahra Kazemi\thanks{Corresponding author: zahra.kazemi@mail.utoronto.ca}}
\author[1]{Craig A. Steeves}

\affil[1]{University of Toronto Institute for Aerospace Studies, 4925 Dufferin Street, Toronto, ON M3H 5T6, Canada}

\begin{document}

\maketitle

\begin{abstract}

The layering approach utilized in fused filament fabrication (FFF) enables the creation of complex designs generated by topology optimization. However, defects associated with the layer-by-layer process, often observed along the thin fusion regions, introduce considerable random variability to the local elastic modulus of the print. The elastic modulus along the fusion layers connecting bulk materials differs from that of the bulk areas. Accurate quantitative measurements of variations in both areas are essential to achieve robust optimized designs. This study aims to quantify the parameters of the random distributions given the surface strain field of the print measured by digital image correlation (DIC). Two statistical properties, mean and standard deviation, are sufficient to characterize the stochastic elastic modulus fields in each region. An efficient neural network model is developed to estimate spatial variations in the local elastic modulus within both bulk and fusion layers. This model is trained on a dataset of synthetic strain fields with known distributions in the elastic modulus fields. It performs well in correlating the elastic modulus with the input strain, as long as the standard deviation is below $60\%$ of the mean of the random field. The predictive accuracy of the model on testing data, measured by the $\mathrm{R^2}$ score, is $0.99$ and $0.95$ for mean and standard deviation in the fusion material. The scores for the bulk material are $0.97$ each. The trained model is implemented to predict the elastic modulus distribution of an FFF-printed material at a print speed of 30 mm/s and an extrusion temperature of 220$^\circ$C, based on its DIC-measured surface strain data. The model predicts a mean and standard deviation of $1.2$ GPa and $1$ GPa for the bulk material and $400$ MPa and $430$ MPa for the fusion region. Validation of these predictions confirms the reliability and credibility of this approach in measuring uncertainty in the local properties of the prints.
\end{abstract}

\keywords{Fused filament fabrication, Local elastic modulus, Uncertainty characterization, Digital image correlation, Machine learning, Robust topology optimization}

\section{Introduction}

Enhancing energy efficiency and reducing the ecological footprint of aviation requires reducing the weight of aircraft components and structures, as lower weight requires less fuel consumption. Structural topology optimization efficiently designs structures using less material while still delivering high performance \cite{bendsoe2013topology, sigmund200199}. Topology optimization, as formulated by Bendsoe and Kikuchi \cite{bendsoe1988generating}, aims to determine the optimal material layout within a design space that maximizes a chosen performance metric, such as structural stiffness, subject to constraints.

The geometric freedom offered by additive manufacturing (AM) enables the fabrication of novel designs with intricate geometric features as are typically created by topology optimization \cite{liu2018current}. Despite this complementary relationship, the inconsistency observed in printed material properties prevents the widespread adoption of additive manufacturing. Process-induced defects, associated with the layer-by-layer process in AM, introduce spatial variations in the local elastic modulus, leading to material property inconsistencies \cite{ahn2002anisotropic, roach2020size, gao2015status, azami2023laser, ngo2018additive, liao2019effect, bellehumeur2004modeling, turner2014review}. Neglecting these uncertainties during the optimization process can lead to unreliable structures. Accurate quantitative measurements of distribution in the elastic modulus field must be integrated at the design stage to achieve robust optimized designs. Robust topology optimization in the presence of uncertainties in material properties and the effect of material variability on structural response and design decisions have been studied \cite{li2023random, pepler2023modelling, asadpoure2011robust}. Broadly, robust topology optimization schemes lead to designs that perform, on average, slightly worse than designs using deterministic material properties, but are much less susceptible to material variability. To achieve this requires statistical information about the distribution of randomly varying material properties; for compliance-minimizing topology optimization, this consists of the mean, variance, and correlation length of the Young’s modulus. Existing robust topology optimization algorithms have been demonstrated using notional material properties: an efficient method for determining the actual statistical distributions of the relevant material properties of an AM material does not exist.

\begin{figure}[htp]
    \centering
    \includegraphics[width=13cm]{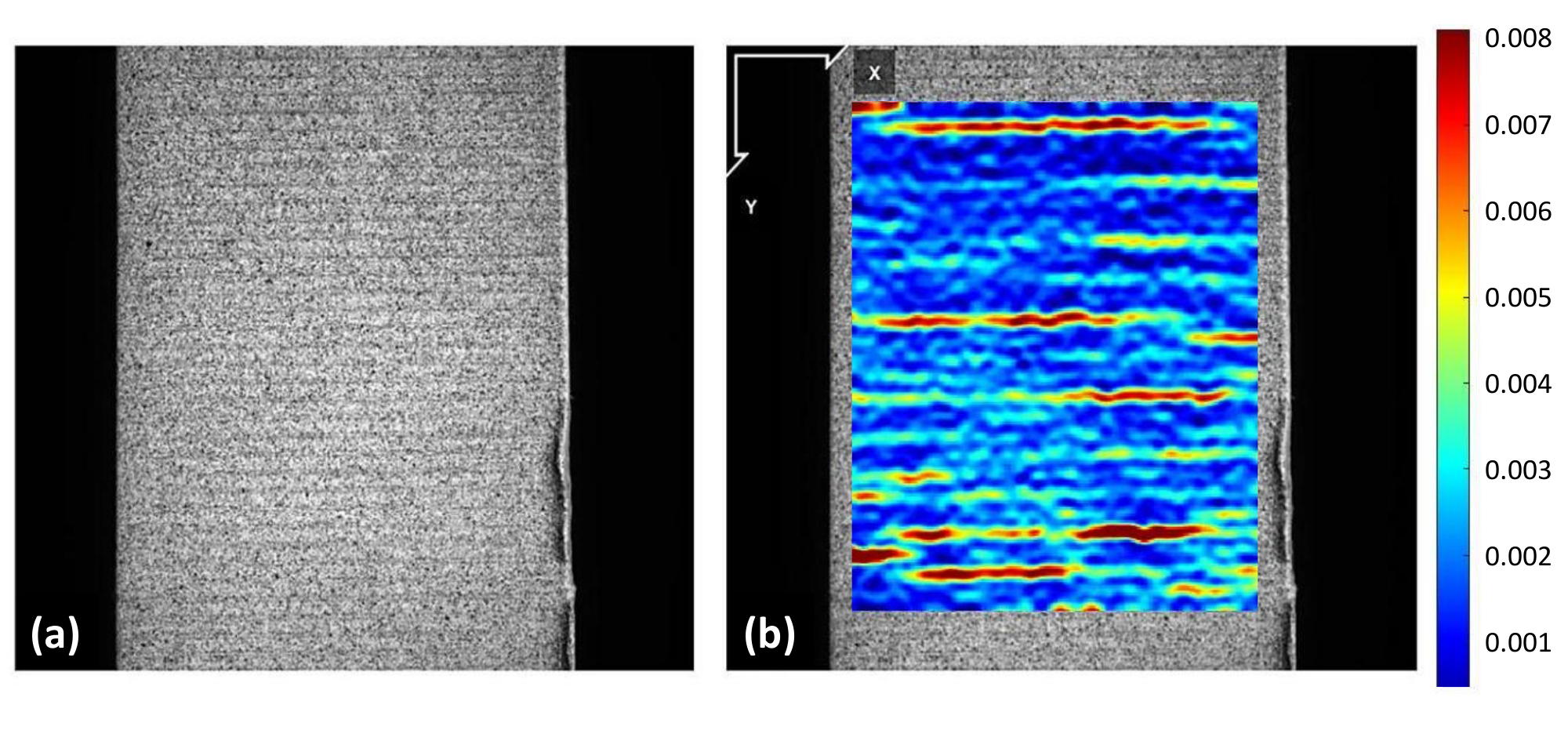}
    \caption{(a) An FFF-printed test specimen speckled with an airbrush and (b) contours of the DIC calculated longitudinal surface strain field, ${\varepsilon_{yy}}$. The measured data reveals horizontally elongated strain features with noticeable variation along the printed path. Notably, the fusion layers display higher strains compared to the bulk layers}
    \label{fig:1}
\end{figure}

The surface strain field of a polylactic acid (PLA) test specimen, fabricated using fused filament fabrication (FFF), is measured via digital image correlation (DIC) \cite{chu1985applications}, as depicted in Figure~\ref{fig:1} (b). DIC analysis of the strained specimen reveals distinct alternating horizontal layers of low and high strains, corresponding to the bulk material and the thin fusion regions connecting layers. FFF-based processes are prone to various defects, such as lack-of-fusion, pores, unmelted materials, and layer misalignments \cite{sanaei2021defects, wang2016plane, lee2014development}, leading to compromised fusion and more compliant interfaces \cite{kazemi2022overall} between adjacent deposited filaments. The goal is to quantify the variation in the local elastic modulus of both regions for the FFF-printed material. Two statistical properties of mean and variance are sufficient to characterize each region's stochastic elastic modulus fields. Conventional mechanical testing techniques are inapplicable for determining the local elastic modulus, as they only measure spatially averaged material properties \cite{abeykoon2020optimization}. Advanced methods like nanoindentation \cite{oliver1992improved, zhang2006characterization} and atomic force microscopy \cite{binnig1986atomic} allow direct measurement of local material properties but are typically limited to very smooth surfaces of the material of interest, whereas printed materials often exhibit significant surface roughness. Given the inability to determine the local elastic modulus directly, we propose an indirect measurement by which variation in the local modulus is inferred from variation in local surface strain during a direct uniaxial tension test.

The printed material contains numerous defects scattered randomly throughout the structure, with overlapping affected areas within the strain field. Developing an accurate and practical model that accounts for the strain dependence between neighboring material points in addressing the underlying stochasticity in the elastic modulus field is required. Establishing an analytical relationship capable of accurately mapping the intricate and highly localized strain field to the elastic modulus field is challenging. We propose a machine learning-based approach aimed at identifying underlying variation in the elastic modulus field and estimating related statistical properties given DIC-calculated surface strain fields. Machine learning technologies are well-established in addressing challenges in various aspects of AM processes, including design process optimization \cite{zhang2018statistical, huang2019surfel, yanamandra2020reverse}, quality control and \textit{in-situ} monitoring \cite{bisheh2021layer, bugatti2022towards}, as well as characterization of effective material properties of the prints \cite{herriott2020predicting, yan2018data, zhan2021novel, nasiri2021machine, zhang2019deep, baturynska2019application}. However, no attempts have been made to address local properties. Building upon the previous study \cite{kazemi2022uncertainty}, this work pioneers the application of machine learning to identify spatial variations in the local elastic modulus of additively manufactured materials and extract the required statistical parameters. The model is trained on a dataset consisting of histograms of synthetic strain fields with known distributions in the bulk and fusion elastic modulus fields. The trained model is able to correlate input strain data with output elastic modulus as long as the standard deviation of the elastic modulus is up to 60\% of the mean. After successful testing with synthetic data, the model is tested with experimental DIC surface strain data of an actual FFF-printed test specimen. Given the histogram of the strain field shown in Figure~\ref{fig:1} (b) as the input data, predictions are made for the elastic modulus distribution parameters for the bulk and fusion materials. The model predictions yield specific values for the mean and standard deviation in the two regions. A method is proposed to validate model predictions, which involves comparing DIC strain measurements with the strain fields predicted by the model. While estimating material variability typically requires extensive experimentation, our model achieves promising predictions based on limited experimental data. Although this study focuses on FFF-printed PLA, the proposed method has the potential for application to other materials and AM techniques, given that appropriate datasets are tailored to the specific material and fabrication process of interest.

\section{Methodology}

\subsection{DIC surface strain measurement}
Full-field displacements and strains can be determined with DIC by tracking the displacement of subsets of physical features on images captured from the test specimen at different times, corresponding to various load and strain conditions, and comparing them to an image of the initial unloaded state. These features are generated by applying a high-contrast black-and-white random speckle to the specimen surface. The speckles evenly cover approximately $50\%$ of the surface. The speckle size is carefully chosen to ensure the required measurement resolution, avoiding both excessively small (less than 3 to 5 pixels) to prevent aliasing and overly large sizes that would necessitate large subsets for tracking, compromising strain calculation resolution. An example of the speckle pattern used in this study is presented in Figure~\ref{fig:1} (a). To create this pattern, the surface of an FFF-printed PLA test specimen is coated with white paint (Rust-Oleum 7751830 High Heat Enamel Spray). Black speckles (Createx Illustration Opaque) are airbrushed at a pressure of 25 psi from a distance of 45 cm using an Iwata Eclipse Airbrush HP-CS Gravity Feed Dual Action. Care is taken to ensure that the distribution of speckles appears sufficiently random and isotropic, maintaining a uniform presence with visually balanced results. The pattern contains speckles whose physical sizes vary between 75 ${\mu}$m and 150 ${\mu}$m (equivalent to 6 and 12 pixels).



  
A PLA sheet is printed in a flat orientation with all filaments deposited in a consistent single direction at a 90 $^\circ$ raster angle. The sheet consists of 17 layers, each with a height of 0.2 mm. The print head speed is set to 30 mm/s, with the temperature at which the semi-solid filament leaves the extrusion nozzle and the build stage temperature set to 220 $^\circ$C and 60 $^\circ$C, respectively. A rectangular tensile specimen of length, width, and thickness 180 mm, 22 mm, and 3.5 mm, respectively, is cut from this sheet. The test specimen, speckled with a similar pattern in Figure~\ref{fig:1} (a), is subjected to a controlled tensile load applied in an MTS load frame. The tests are carried out in displacement control mode, with a piston displacement rate of 0.25 mm/min applied along the axial direction, perpendicular to the printing path. Images of the specimen in its undeformed state and different strained states are taken using a FLIR camera. Analysis of these images, tracking subsets of speckles with a radius of $20$ pixels spaced every $4$ pixels apart, is performed using the open-source 2D Ncorr MATLAB software \cite{blaber2015ncorr}. The subset radius is chosen based on the size of the speckles, ensuring it is large enough to contain at least three speckles \cite{sutton2009image}. To achieve high-resolution local strain measurement in fusion regions, the strain radius is carefully chosen to exclude displacement data from outside the region of interest. With a fusion layer thickness of around 130 ${\mu}$m and image pixel size of 12.5 ${\mu}$m, the strain radius is set to $5$ pixels. Anisotropic mechanical properties are commonly observed in printed materials. This study focuses on reporting axial strains as it is the only direction that exhibits significant variation in strain, as expected given the direction of the load. Figure~\ref{fig:1} illustrates the axial surface strain field of the PLA FFF-printed specimen measured by DIC.

\subsection{Dataset preparation for training machine learning model}
\begin{figure}[t]
\centering\includegraphics[width=0.45\linewidth]{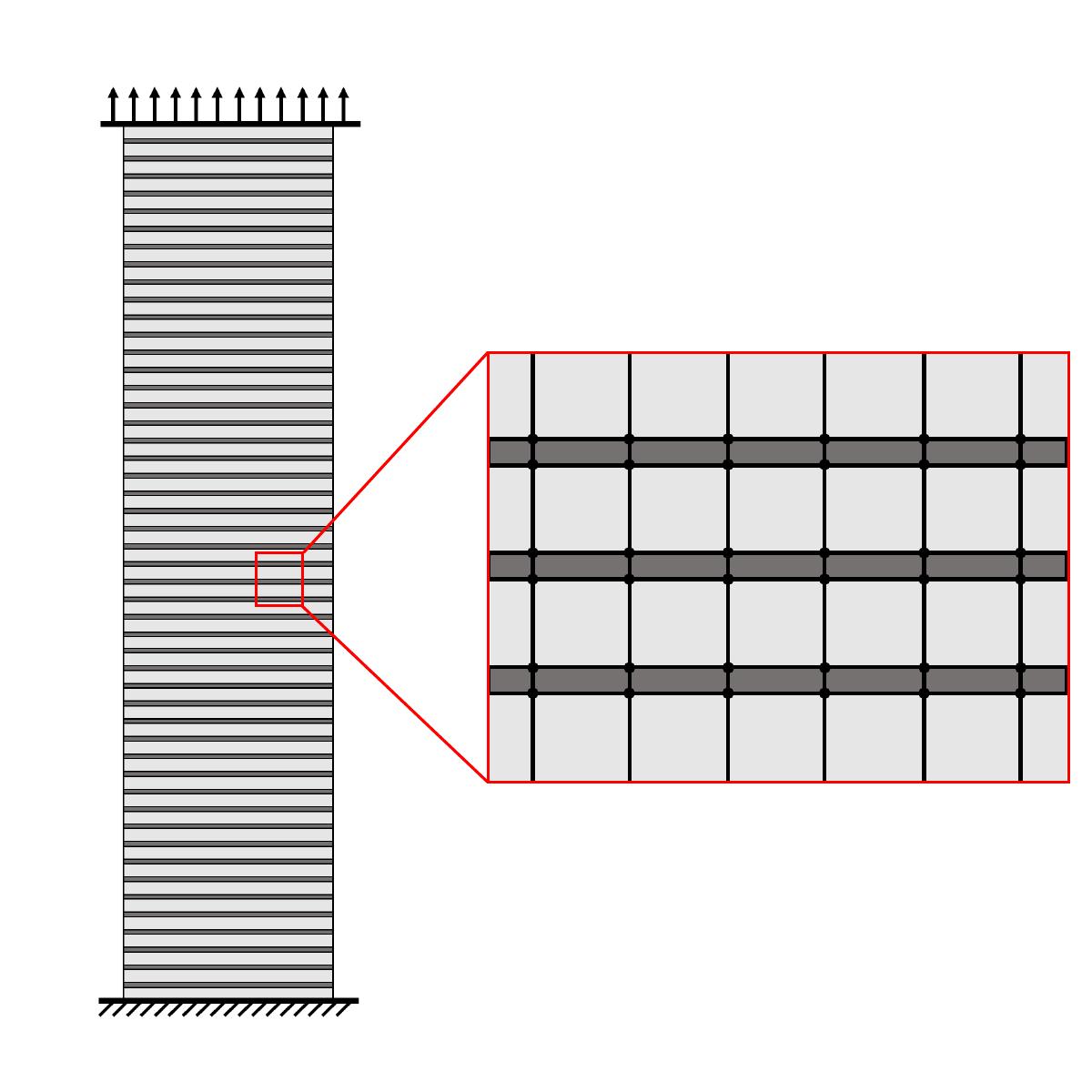}
\caption{A representative finite element mesh used to model printed material. Light grey indicates the bulk material, while dark grey represents the fusion layers.}\label{fig:5}
\end{figure}

Figure~\ref{fig:1} (b) illustrates the distinct horizontal layering of strain in alternating bulk and fusion materials characterized by different elastic modulus distributions. The objective is to train a machine learning model capable of estimating the statistical properties of mean and standard deviation in bulk and fusion elastic modulus random fields independently, based on the DIC surface strain data. The mean and standard deviation are the key statistical properties of the distribution of Young’s modulus necessary for robust topology optimization. While the correlation length is also a desirable parameter, it is very difficult to discern small differences in correlation length using the approach described here. The model is trained on a dataset consisting of histograms of synthetic strain fields with known bulk and fusion elastic modulus distributions. The strain fields are generated through finite element analysis (FEA) of synthetic elastic modulus random fields with predefined means and standard deviations for bulk and fusion layers. Optical microscope measurements indicate that the mean thickness of the fusion layer is approximately 130 ${\mu}$m, while that of the bulk layer is around 400 ${\mu}$m. The 1:3 ratio is maintained when constructing the mesh for FEA of the elastic modulus fields and determining the element sizes in the bulk layers. For discretization, each bulk layer is represented by a single row of linear square elements, and each fusion layer is modeled with a single row of linear rectangular elements with an aspect ratio of 3. The structure of the finite element mesh is given in Figure~\ref{fig:5}. With the construction of 50 groups of bulk and fusion layers, a total of $17,000$ linear elements are used to conduct the FE simulation of the problem.

Layers within the same region, whether bulk or fusion, may not exhibit identical distributions (see Figure~\ref{fig:1} (b)). While independently generating one-dimensional elastic modulus fields results in identical distributions across different layers within the same region, introducing variations in distributions among these layers requires a different approach. This involves generating a two-dimensional elastic modulus field for the entire fusion region using discrete Karhunen-Loève (KL) expansion with a predefined mean and standard deviation, along with a correlation length of 1 mm. Horizontal sequences of values are then extracted from this field. A similar procedure is applied to the bulk region. These extracted fields are then assembled, alternating fusion and bulk layers, to constrict a synthetic elastic modulus field for the printed material with alternating bulk and fusion layers. Five hundred realizations of the elastic modulus fields are generated for each combination of means and standard deviations.

The KL expansion can be regarded as an abstract discretization of a random process using a series expansion of a set of orthogonal random variables based on the spectral decomposition of the underlying correlation kernel. Any realization of the elastic modulus random field $\textit{E}(\textit{\textbf{x}}, \omega)$ is approximated with the discrete KL expansion as a truncated sum of $\textit{N}$ elements.The spatial domain of a continuous random field is divided into discrete elements. The random variable at an element center is used to represent the randomness at any point within that element. These elements are often chosen to be consistent with those used in structural FEA. Accordingly, the continuous random field $\textit{E}(\omega)$ is a random vector. It is defined as:

\begin{equation}
\textit{E}(\omega) = 
\begin{bmatrix}
\textit{E}(\textit{\textbf{X}}_1, \omega) \\
\textit{E}(\textit{\textbf{X}}_2, \omega) \\
\vdots   \\
\textit{E}(\textit{\textbf{X}}_{n_{ele}}, \omega)
\end{bmatrix}
,
\end{equation}

\noindent where $\textit{\textbf{X}}_i$ is the position vector of the centroid of the $\textit{i}$th element. The discrete correlation matrix is given by:

\begin{equation}
\mathrm{\textbf{P}} = 
\begin{bmatrix}
\rho(\textit{\textbf{X}}_1, \textit{\textbf{X}}_1) & \rho(\textit{\textbf{X}}_1, \textit{\textbf{X}}_2) & \cdots & \rho(\textit{\textbf{X}}_1, \textit{\textbf{X}}_{n_{ele}})\\
\rho(\textit{\textbf{X}}_2, \textit{\textbf{X}}_1) & \rho(\textit{\textbf{X}}_2, \textit{\textbf{X}}_2) & \cdots & \rho(\textit{\textbf{X}}_2, \textit{\textbf{X}}_{n_{ele}})\\
\vdots  & \vdots & \cdots &  \vdots\\
\rho(\textit{\textbf{X}}_{n_{ele}}, \textit{\textbf{X}}_1) & \rho(\textit{\textbf{X}}_{n_{ele}}, \textit{\textbf{X}}_2) & \cdots & \rho(\textit{\textbf{X}}_{n_{ele}}, \textit{\textbf{X}}_{n_{ele}})
\end{bmatrix}
,
\end{equation}

\noindent The elements of this matrix are determined using the exponential square correlation function $\rho(\textit{\textbf{X}},\textit{\textbf{X}}^\prime)$, which is given by:

\begin{equation}\label{eqn:1}
\rho(\textit{\textbf{X}},\textit{\textbf{X}}^\prime) = \sigma^2 e^{(-\frac{|\textit{\textbf{X}}-\textit{\textbf{X}}^\prime|^2}{\textit{l}^2})},
\end{equation}

\noindent where $\textit{l}$ is the correlation length. Using the eigenpairs ($\lambda_i$, $\phi_{i}$) of the correlation matrix $\mathrm{\textbf{P}}$, the random vector $\textit{E}(\omega)$ is approximated by the discrete KL expansion:

\begin{equation}\label{eqn:2}
\textit{E}(\omega) = \mu + \sum_{i=1}^{N} \sqrt{\lambda_i} \phi_{i} \xi_i (\omega).
\end{equation}

The parameters $\xi_i (\omega)$ are independent and identically distributed (i.i.d) standard Gaussian random variables with $\omega$ representing a stochastic realization. $\mu$ and $\sigma$ denote the mean value and standard deviation of the elastic modulus random field, respectively. The order of expansion $\textit{N}$ indicates the number of eigenvalues and eigenvectors used in the truncated approximation to describe the random field. To represent the elastic modulus, a lognormal random field is used, chosen for its adherence to positivity, which is a desirable attribute for representing modulus fields as they are strictly positive. The exponential of a truncated KL expansion is used to describe the lognormal random field. Additionally, there may be inaccuracies associated with KL in representing random elastic modulus fields, particularly regarding the statistical properties of the approximated fields. Discrepancies may arise between the defined statistical properties of mean and standard deviation for the KLE and the mean and variance of generated random fields. The discrete version of the KL expansion can help mitigate such inaccuracies.

\begin{figure}
\centering\includegraphics[width=0.9\linewidth]{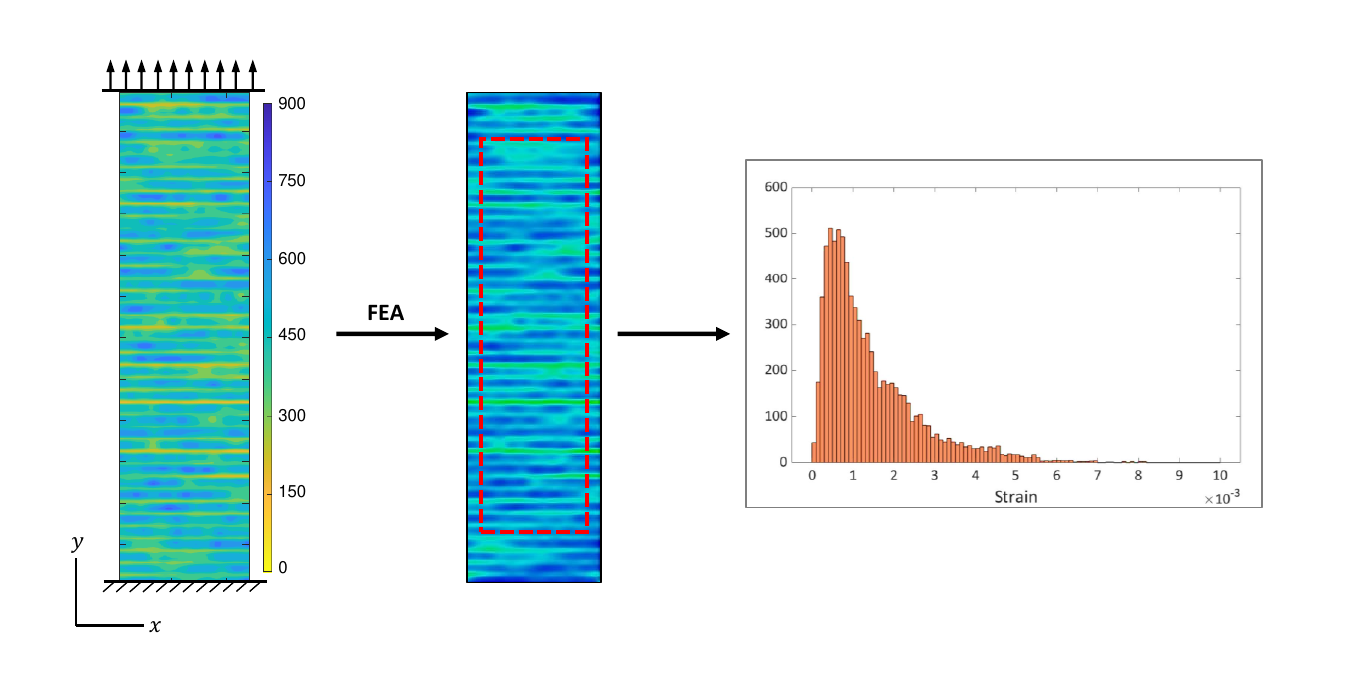}
\caption{The left figure displays a synthetic elastic modulus field with applied boundary conditions. The alternating 1D layers of high and low modulus fields represent the bulk material and the thin fusion layers. The middle figure is the corresponding axial strain field, $\varepsilon_{yy}$, and the right figure shows a histogram of the strain data from the region of interest (highlighted by the dotted red box) used for training the model.}\label{fig:3}
\end{figure}

The ensemble of one-dimensional fields, derived from two-dimensional realizations of bulk and fusion elastic modulus random fields generated by discrete KL expansion, models the elastic modulus field for the printed material (see Figure~\ref{fig:3}). The meshed structure is subject to axial displacement along the upper edge and remains fixed at the lower boundary, as shown in Figure~\ref{fig:3}, resembling the conditions experienced by a test specimen during tensile testing for DIC calculations to collect surface strain data. The boundary effect near the top and bottom edges is significant; thus, only strain values within a rectangular domain in the middle are considered for training purposes. Histograms of the axial strain fields ($\varepsilon_{yy}$), rather than the fields themselves, are given to the machine learning model as the input features as they have lower dimensionality while retaining sufficient information about the random fields. Hence, the dataset comprises histograms of synthetic strain fields (features) with known statistical properties of means and standard deviations in bulk and fusion elastic modulus fields (targets). The dataset consists of 12,000 instances, including 24 different combinations of means and standard deviations for the elastic modulus fields, with 500 realizations for each combination. The data is rescaled using the RobustScaler technique \cite{ahsan2021effect} to ensure that all features are on a common scale, and features with minimal variance are dropped. A neural network with three hidden layers of 150 neurons, along with a dropout layer having a 0.3 rate, demonstrated superior predictive accuracy. The model is trained on a set of 10,000 training data and subsequently evaluated on 2,000 testing data, selected randomly from the entire dataset pool. The model is trained with the Adam optimizer \cite{kingma2014adam}.

\section{Results and discussion}

\begin{figure}
\centering\includegraphics[width=0.75\linewidth]{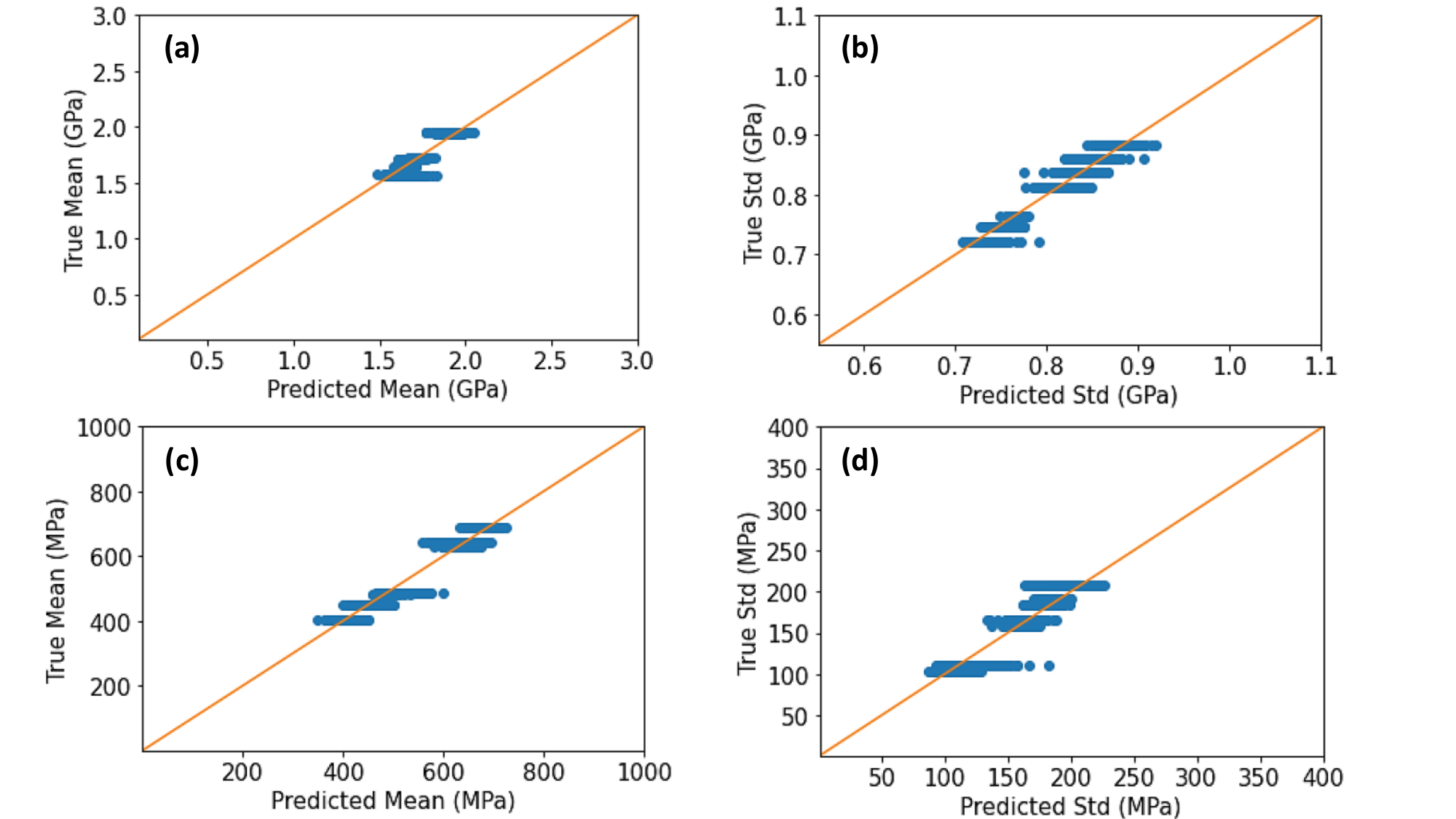}
\caption{Comparison of the predicted target values and true target values for (a) mean and (b) standard deviation in bulk Young’s modulus fields, and (c) mean and (d) standard deviation in fusion Young’s modulus field. The slope of the orange line is 1}\label{fig:4}
\end{figure}

Table~\ref{tab:1} presents the $\mathrm{R^2}$ scores of the trained model on the testing data for four targets. The proximity of $\mathrm{R^2}$ scores to one for all targets demonstrates that the model can accurately estimate parameters of means and standard deviations, thereby capturing the underlying distribution within the elastic modulus random fields given the intricate strain field. Figure~\ref{fig:4} is a visualization of the predictive performance of the model for four targets on the test data. These plots compare the actual targets against the predictions made by the trained model for mean and standard deviation values in both bulk and fusion elastic modulus fields. 

\begin{table}
\centering
\caption{$\mathrm{R^2}$ scores of the trained neural network on the testing data for individual targets}  
{\begin{tabular}{l l l}
\toprule
\textbf{Region} & \textbf{Statistical parameter} & \textbf{$\mathrm{R^2}$} \\
\midrule
Bulk     & Mean               &  0.97 \\
         & Standard deviation &  0.97 \\
\midrule         
Fusion   & Mean               &  0.99 \\
         & Standard deviation &  0.95 \\
\midrule
\end{tabular}}
\label{tab:1}
\end{table}

The effectiveness of the model in detecting underlying patterns in the elastic modulus depends on the magnitude of the variations within the bulk and fusion elastic modulus fields compared to the mean values. As long as the standard deviation remains below 60\% of the mean values of these fields, the model performs well in correlating the elastic modulus with the input strain. In such cases, the model can successfully identify relationships between the two variables. For higher standard deviations in the local elastic modulus, the performance of the model is compromised, and the model fails to establish a correlation between the target elastic modulus and input strain. This is likely because a large variance in the modulus in the bulk or fusion regions influences the strain field in the other region, leading to a degradation in the correlation between the strain and the elastic modulus in the affected region. High variances can obscure any changes in the mean and standard deviation of the elastic modulus in the corresponding strain fields, making it challenging to recognize their corresponding strain fields and make accurate predictions about the elastic modulus.

The model demonstrates satisfactory performance when dealing with synthetic strain data, successfully predicting the underlying variations in the local elastic modulus for both the bulk and fusion areas, provided the variances are not excessive. These results indicate the model captures correlations between strain and elastic modulus in synthetic settings. Following successful testing using synthetic data, the trained model is tested with experimental DIC surface strain data of an actual FFF-printed test specimen. The histogram of the strain field shown in Figure~\ref{fig:1} (b) is imported as the input data, and predictions are made for the four specified targets that represent elastic modulus distributions for the bulk and fusion materials in the printed specimen. The model predictions yield specific values for the mean and variance in the two regions. The model predicts a mean and standard deviation of 1.2 GPa and 1 GPa for the bulk material. Similarly, for the fusion region, the model predicts a mean and standard deviation of 400 MPa and 430 MPa. These values of the mean and standard deviation are consistent with the ranges of these properties studied by Li and Steeves \cite{li2023random} for use in robust topology optimization.

A method to validate model predictions involves comparing DIC strain measurements with the strain fields obtained from finite element calculations on meshes populated by modulus fields with the same means and standard deviations as predicted by the model. Two-dimensional bulk and fusion modulus fields are generated using discrete KL expansion with model-predicted means and standard deviations. One-dimensional fields, taken from the two-dimensional modulus fields, are alternately assembled to construct a complete elastic modulus field. Strain fields corresponding to these modulus fields are then calculated through FEA with boundary conditions identical to those experienced by the test specimen during DIC calculation (Figure~\ref{fig:3} (a)). The histogram of axial strains is then compared with experimental DIC strain data, with separate analyses for bulk and fusion strains. The quantitative results are summarized in Table~\ref{tab:2}. 

\begin{table}[htp]
\centering
\caption{Comparison of the distribution in the DIC surface strain field with the predicted strain field for bulk and fusion regions}  
{\begin{tabular}{l l l l l}
\toprule
\textbf{Region} & \textbf{Statistical parameter} & \textbf{DIC strain} & \textbf{Predicted strain} & \textbf{Error (\%)} \\
\midrule
Bulk     & Mean               &  8.2e-04 & 8.3e-04 &  0.7 \\
         & Standard deviation &  6.2e-04 & 5.8e-04 &  7   \\
\midrule         
Fusion   & Mean               &  1.6e-03 & 2.0e-03 &  24  \\
         & Standard deviation &  1.3e-03 & 1.2e-03 &  10  \\
\midrule
\end{tabular}}
\label{tab:2}
\end{table}

In the fusion region, the standard deviation of the DIC-measured strain field is 10\% larger than that of the field obtained from predictions. The mean of the DIC strain field is 24\% smaller than the corresponding prediction. This discrepancy primarily arises from the presence of negative strain values in the DIC strain field, which contribute to a larger standard deviation and a lower mean. Similar observations apply to strain values in the bulk region, primarily due to the presence of negative strains within the DIC data. The negative strains primarily are attributed to the inherent noise in the DIC strain calculations, particularly notable for small strains and minor deformations of the printed specimen. Despite these discrepancies, the overall agreement between the model predictions and the experiments confirms the feasibility of this approach in measuring variation in the local elastic modulus of the prints.

\section{Conclusion}
The main contribution of this research is the development of a method for extracting information about the underlying stochastic properties of highly complicated, dense strain field data. This is achieved by considering the strain dependencies among neighboring material points. A neural network model capable of identifying variations in local elastic properties within both fusion and bulk materials is developed, provided that the standard deviations remain below 60\% of the mean values within the random fields. The model performance is compromised for higher standard deviations, failing to establish a meaningful correlation between the target elastic modulus and input strain data. On synthetic testing data, high $\mathrm{R}^2$ scores of $0.99$ and $0.95$ for mean and standard deviation in the fusion material, respectively, are achieved. For the bulk material, the scores are $0.97$ for both mean and standard deviation. The trained model is implemented to infer the elastic modulus distribution of an FFF-printed material at specific print process parameters. At a print head speed of $30$ mm/s, with the semi-solid filament leaving the extrusion nozzle at 220$^\circ$C and the build stage temperature set at 60$^\circ$C, the model predicts a mean and standard deviation of $1.2$ GPa and $1$ GPa for the bulk material, and $400$ MPa and $430$ MPa for the fusion region, based on DIC-calculated surface strain data. The model predictive accuracy is validated through a comparative analysis between the DIC strain data and the strain field obtained from model predictions. The comparison shows that the difference between the means of the DIC strain field and the predicted strain field is 24\%, while the difference between standard deviations is 10\% in the fusion regions. For the bulk material, the difference between means is less than 1\%, and the difference between standard deviations is 7\%. These discrepancies are mainly due to negative strain values in the DIC strain field, leading to a larger standard deviation and a lower mean in DIC-measured strain fields. The comparison results confirm the feasibility of this approach in measuring local elastic properties given high-resolution DIC surface strain fields. 

While estimating material variability commonly requires extensive experimentation, our model provides predictions based on very limited experimental data. Although this novel approach is specifically designed for FFF processes, it shows promise for extension to alternative AM methods and materials. To enable such a broader application, appropriate datasets tailored to the specific material and fabrication process of interest must be generated for model training. Moving forward, integrating the measured variations at the design stage will result in robust and conservative designs capable of withstanding uncertainties in material properties arising from the AM layering process. This bridges the gap between topology optimization and AM, enabling the convenient fabrication of complex configurations that would face manufacturability restrictions using conventional techniques while accounting for uncertainties at the design stage.

\section*{Acknowledgments}
The authors extend their appreciation to Autodesk and the National Sciences and Engineering Research Council of Canada (NSERC) for funding and supporting this research project.

\section*{Declarations}
On behalf of all authors, the corresponding author states that there is no conflict of interest.


\printbibliography 

@book{bendsoe2013topology,
  title={Topology optimization: theory, methods, and applications},
  author={Bendsoe, Martin Philip and Sigmund, Ole},
  year={2013},
  publisher={Springer Science \& Business Media}
}

@article{sigmund200199,
  title={A 99 line topology optimization code written in Matlab},
  author={Sigmund, Ole},
  journal={Structural and multidisciplinary optimization},
  volume={21},
  pages={120--127},
  year={2001},
  publisher={Springer}
}

@article{bendsoe1988generating,
  title={Generating optimal topologies in structural design using a homogenization method},
  author={Bends{\o}e, Martin Philip and Kikuchi, Noboru},
  journal={Computer methods in applied mechanics and engineering},
  volume={71},
  number={2},
  pages={197--224},
  year={1988},
  publisher={Elsevier}
}

@article{liu2018current,
  title={Current and future trends in topology optimization for additive manufacturing},
  author={Liu, Jikai and Gaynor, Andrew T and Chen, Shikui and Kang, Zhan and Suresh, Krishnan and Takezawa, Akihiro and Li, Lei and Kato, Junji and Tang, Jinyuan and Wang, Charlie CL and others},
  journal={Structural and multidisciplinary optimization},
  volume={57},
  number={6},
  pages={2457--2483},
  year={2018},
  publisher={Springer}
}

@article{sanaei2021defects,
  title={Defects in additive manufactured metals and their effect on fatigue performance: A state-of-the-art review},
  author={Sanaei, Niloofar and Fatemi, Ali},
  journal={Progress in Materials Science},
  volume={117},
  pages={100724},
  year={2021},
  publisher={Elsevier}
}

@article{roach2020size,
  title={Size-dependent stochastic tensile properties in additively manufactured 316L stainless steel},
  author={Roach, Ashley M and White, Benjamin C and Garland, Anthony and Jared, Bradley H and Carroll, Jay D and Boyce, Brad L},
  journal={Additive Manufacturing},
  volume={32},
  pages={101090},
  year={2020},
  publisher={Elsevier}
}

@article{gao2015status,
  title={The status, challenges, and future of additive manufacturing in engineering},
  author={Gao, Wei and Zhang, Yunbo and Ramanujan, Devarajan and Ramani, Karthik and Chen, Yong and Williams, Christopher B and Wang, Charlie CL and Shin, Yung C and Zhang, Song and Zavattieri, Pablo D},
  journal={Computer-aided design},
  volume={69},
  pages={65--89},
  year={2015},
  publisher={Elsevier}
}

@article{ngo2018additive,
  title={Additive manufacturing (3D printing): A review of materials, methods, applications and challenges},
  author={Ngo, Tuan D and Kashani, Alireza and Imbalzano, Gabriele and Nguyen, Kate TQ and Hui, David},
  journal={Composites Part B: Engineering},
  volume={143},
  pages={172--196},
  year={2018},
  publisher={Elsevier}
}

@article{liao2019effect,
  title={Effect of porosity and crystallinity on 3D printed PLA properties},
  author={Liao, Yuhan and Liu, Chang and Coppola, Bartolomeo and Barra, Giuseppina and Di Maio, Luciano and Incarnato, Loredana and Lafdi, Khalid},
  journal={Polymers},
  volume={11},
  number={9},
  pages={1487},
  year={2019},
  publisher={MDPI}
}

@article{bellehumeur2004modeling,
  title={Modeling of bond formation between polymer filaments in the fused deposition modeling process},
  author={Bellehumeur, C{\'e}line and Li, Longmei and Sun, Qian and Gu, Peihua},
  journal={Journal of manufacturing processes},
  volume={6},
  number={2},
  pages={170--178},
  year={2004},
  publisher={Elsevier}
}

@article{turner2014review,
  title={A review of melt extrusion additive manufacturing processes: I. Process design and modeling},
  author={Turner, Brian N and Strong, Robert and Gold, Scott A},
  journal={Rapid prototyping journal},
  volume={20},
  number={3},
  pages={192--204},
  year={2014},
  publisher={Emerald Group Publishing Limited}
}

@article{ahn2002anisotropic,
  title={Anisotropic material properties of fused deposition modeling ABS},
  author={Ahn, Sung-Hoon and Montero, Michael and Odell, Dan and Roundy, Shad and Wright, Paul K},
  journal={Rapid prototyping journal},
  volume={8},
  number={4},
  pages={248--257},
  year={2002},
  publisher={MCB UP Ltd}
}

@article{chu1985applications,
  title={Applications of digital-image-correlation techniques to experimental mechanics},
  author={Chu, TC and Ranson, WF and Sutton, Michael A},
  journal={Experimental mechanics},
  volume={25},
  pages={232--244},
  year={1985},
  publisher={Kluwer Academic Publishers}
}

@article{wang2016plane,
  title={In-plane shape-deviation modeling and compensation for fused deposition modeling processes},
  author={Wang, Andi and Song, Suoyuan and Huang, Qiang and Tsung, Fugee},
  journal={IEEE Transactions on Automation Science and Engineering},
  volume={14},
  number={2},
  pages={968--976},
  year={2016},
  publisher={IEEE}
}

@article{lee2014development,
  title={Development of a hybrid rapid prototyping system using low-cost fused deposition modeling and five-axis machining},
  author={Lee, Wei-chen and Wei, Ching-chih and Chung, Shan-Chen},
  journal={Journal of Materials Processing Technology},
  volume={214},
  number={11},
  pages={2366--2374},
  year={2014},
  publisher={Elsevier}
}

@article{oliver1992improved,
  title={An improved technique for determining hardness and elastic modulus using load and displacement sensing indentation experiments},
  author={Oliver, Warren Carl and Pharr, George Mathews},
  journal={Journal of materials research},
  volume={7},
  number={6},
  pages={1564--1583},
  year={1992},
  publisher={Cambridge University Press}
}

@article{zhang2006characterization,
  title={Characterization of mechanical properties of polymers by nanoindentation tests},
  author={Zhang, CY and Zhang, YW and Zeng, KY and Shen, L},
  journal={Philosophical Magazine},
  volume={86},
  number={28},
  pages={4487--4506},
  year={2006},
  publisher={Taylor \& Francis}
}

@article{binnig1986atomic,
  title={Atomic force microscope},
  author={Binnig, Gerd and Quate, Calvin F and Gerber, Ch},
  journal={Physical review letters},
  volume={56},
  number={9},
  pages={930},
  year={1986},
  publisher={APS}
}

@article{abeykoon2020optimization,
  title={Optimization of fused deposition modeling parameters for improved PLA and ABS 3D printed structures},
  author={Abeykoon, Chamil and Sri-Amphorn, Pimpisut and Fernando, Anura},
  journal={International Journal of Lightweight Materials and Manufacture},
  volume={3},
  number={3},
  pages={284--297},
  year={2020},
  publisher={Elsevier}
}

@article{zhang2018statistical,
  title={A statistical method for build orientation determination in additive manufacturing},
  author={Zhang, Yicha and Harik, Ramy and Fadel, Georges and Bernard, Alain},
  journal={Rapid Prototyping Journal},
  volume={25},
  number={1},
  pages={187--207},
  year={2018},
  publisher={Emerald Publishing Limited}
}

@article{huang2019surfel,
  title={Surfel convolutional neural network for support detection in additive manufacturing},
  author={Huang, Jida and Kwok, Tsz-Ho and Zhou, Chi and Xu, Wenyao},
  journal={The International Journal of Advanced Manufacturing Technology},
  volume={105},
  pages={3593--3604},
  year={2019},
  publisher={Springer}
}

@article{yanamandra2020reverse,
  title={Reverse engineering of additive manufactured composite part by toolpath reconstruction using imaging and machine learning},
  author={Yanamandra, Kaushik and Chen, Guan Lin and Xu, Xianbo and Mac, Gary and Gupta, Nikhil},
  journal={Composites Science and Technology},
  volume={198},
  pages={108318},
  year={2020},
  publisher={Elsevier}
}

@article{bisheh2021layer,
  title={A layer-by-layer quality monitoring framework for 3D printing},
  author={Bisheh, Mohammad Najjartabar and Chang, Shing I and Lei, Shuting},
  journal={Computers \& Industrial Engineering},
  volume={157},
  pages={107314},
  year={2021},
  publisher={Elsevier}
}

@article{bugatti2022towards,
  title={Towards real-time in-situ monitoring of hot-spot defects in L-PBF: a new classification-based method for fast video-imaging data analysis},
  author={Bugatti, Matteo and Colosimo, Bianca Maria},
  journal={Journal of Intelligent Manufacturing},
  volume={33},
  number={1},
  pages={293--309},
  year={2022},
  publisher={Springer}
}

@inproceedings{kazemi2022uncertainty,
  title={Uncertainty quantification in material properties of additively manufactured materials for application in topology optimization},
  author={Kazemi, Zahra and Steeves, Craig A},
  booktitle={ASME international mechanical engineering congress and exposition},
  volume={86656},
  pages={V003T04A009},
  year={2022},
  organization={American Society of Mechanical Engineers}
}

@article{herriott2020predicting,
  title={Predicting microstructure-dependent mechanical properties in additively manufactured metals with machine-and deep-learning methods},
  author={Herriott, Carl and Spear, Ashley D},
  journal={Computational Materials Science},
  volume={175},
  pages={109599},
  year={2020},
  publisher={Elsevier}
}

@article{yan2018data,
  title={Data-driven prediction of mechanical properties in support of rapid certification of additively manufactured alloys},
  author={Yan, Fuyao and Chan, Yu-Chin and Saboo, Abhinav and Shah, Jiten and Olson, Gregory B and Chen, Wei},
  journal={Computer Modeling in Engineering \& Sciences},
  volume={117},
  number={3},
  pages={343--366},
  year={2018},
  publisher={Tech Science Press}
}

@article{zhan2021novel,
  title={A novel approach based on the elastoplastic fatigue damage and machine learning models for life prediction of aerospace alloy parts fabricated by additive manufacturing},
  author={Zhan, Zhixin and Li, Hua},
  journal={International Journal of Fatigue},
  volume={145},
  pages={106089},
  year={2021},
  publisher={Elsevier}
}

@article{nasiri2021machine,
  title={Machine learning in predicting mechanical behavior of additively manufactured parts},
  author={Nasiri, Sara and Khosravani, Mohammad Reza},
  journal={Journal of materials research and technology},
  volume={14},
  pages={1137--1153},
  year={2021},
  publisher={Elsevier}
}

@article{zhang2019deep,
  title={Deep learning-based tensile strength prediction in fused deposition modeling},
  author={Zhang, Jianjing and Wang, Peng and Gao, Robert X},
  journal={Computers in industry},
  volume={107},
  pages={11--21},
  year={2019},
  publisher={Elsevier}
}

@article{baturynska2019application,
  title={Application of machine learning techniques to predict the mechanical properties of polyamide 2200 (PA12) in additive manufacturing},
  author={Baturynska, Ivanna},
  journal={Applied Sciences},
  volume={9},
  number={6},
  pages={1060},
  year={2019},
  publisher={MDPI}
}

@article{blaber2015ncorr,
  title={Ncorr: open-source 2D digital image correlation matlab software},
  author={Blaber, Justin and Adair, Benjamin and Antoniou, Antonia},
  journal={Experimental Mechanics},
  volume={55},
  number={6},
  pages={1105--1122},
  year={2015},
  publisher={Springer}
}

@book{sutton2009image,
  title={Image correlation for shape, motion and deformation measurements: basic concepts, theory and applications},
  author={Sutton, Michael A and Orteu, Jean Jose and Schreier, Hubert},
  year={2009},
  publisher={Springer Science \& Business Media}
}

@article{ahsan2021effect,
  title={Effect of data scaling methods on machine learning algorithms and model performance},
  author={Ahsan, Md Manjurul and Mahmud, MA Parvez and Saha, Pritom Kumar and Gupta, Kishor Datta and Siddique, Zahed},
  journal={Technologies},
  volume={9},
  number={3},
  pages={52},
  year={2021},
  publisher={MDPI}
}

@article{kingma2014adam,
  title={Adam: A method for stochastic optimization},
  author={Kingma, Diederik P and Ba, Jimmy},
  journal={arXiv preprint arXiv:1412.6980},
  year={2014}
}

@article{li2023random,
  title={Random matrix theory for robust topology optimization with material uncertainty},
  author={Li, Linxi and Steeves, Craig A},
  journal={Structural and Multidisciplinary Optimization},
  volume={66},
  number={11},
  pages={240},
  year={2023},
  publisher={Springer}
}

@phdthesis{pepler2023modelling,
  title={Modelling Additively Manufactured Material for Applications in Topology Optimization},
  author={Pepler, Daniel},
  year={2023},
  school={University of Toronto (Canada)}
}

@article{asadpoure2011robust,
  title={Robust topology optimization of structures with uncertainties in stiffness--Application to truss structures},
  author={Asadpoure, Alireza and Tootkaboni, Mazdak and Guest, James K},
  journal={Computers \& Structures},
  volume={89},
  number={11-12},
  pages={1131--1141},
  year={2011},
  publisher={Elsevier}
}

@article{azami2023laser,
  title={Laser powder bed fusion of Alumina/Fe--Ni ceramic matrix particulate composites impregnated with a polymeric resin},
  author={Azami, Mohammad and Siahsarani, Armin and Hadian, Amir and Kazemi, Zahra and Rahmatabadi, Davood and Kashani-Bozorg, Seyed Farshid and Abrinia, Karen},
  journal={Journal of Materials Research and Technology},
  volume={24},
  pages={3133--3144},
  year={2023},
  publisher={Elsevier}
}

@inproceedings{kazemi2022overall,
  title={Overall Mechanical Properties of Self-Healing Composites: Effects of Microcapsules Shape, Volume Concentration, Shell Thickness, and Material Properties},
  author={Kazemi, Zahra and Azami, Mohammad},
  booktitle={ASME International Mechanical Engineering Congress and Exposition},
  volume={86717},
  pages={V009T12A001},
  year={2022},
  organization={American Society of Mechanical Engineers}
}


\end{document}